# Current-induced viscoelastic topological unwinding of metastable skyrmion strings

Fumitaka Kagawa[1,†], Hiroshi Oike[1], Wataru Koshibae[1], Akiko Kikkawa[1], Yoshihiro Okamura[2], Yasujiro Taguchi[1], Naoto Nagaosa[1,2] & Yoshinori Tokura[1,2]

[1] *RIKEN Center for Emergent Matter Science (CEMS), Wako 351-0198, Japan*

[2] *Department of Applied Physics, The University of Tokyo, Tokyo 113-8656, Japan*

† To whom correspondence should be addressed. E-mail: fumitaka.kagawa@riken.jp

Abstract

**In the MnSi bulk chiral magnet, magnetic skyrmion strings of 17 nm in diameter appear in the form of a lattice, penetrating the sample thickness, 10–1,000 μm. Although such a bundle of skyrmion strings may exhibit complex soft-matter-like dynamics when starting to move under the influence of a random pinning potential, the details remain highly elusive. Here, we show that a metastable skyrmion-string lattice is subject to topological unwinding under the application of pulsed currents of $3\text{–}5\times10^6$ A m$^{-2}$ rather than being transported, as evidenced by measurements of the topological Hall effect. The critical current density above which the topological unwinding occurs is larger for a shorter pulse width, reminiscent of the viscoelastic characteristics accompanying the pinning-creep transition observed in domain-wall motion. Numerical simulations reveal that current-induced depinning of already segmented skyrmion strings initiates the topological unwinding. Thus, the skyrmion-string length is an element to consider when studying current-induced motion.**

An ordered solid exhibits elastic (reversible) or plastic (irreversible) deformations, depending on the strength of the force applied. Conceptually similar phenomena have also been observed in electronic systems, such as ferroelectric/ferromagnetic domain walls[1,2], flux-line lattices in type-II superconductors[3], and charge/spin density waves[4,5]. Domain walls, for example, are more or less meandering by nature in real materials because of random pinning-potential, and they further deform under an effective force, such as a magnetic/electric field or electric current. When the effective force is relatively weak, the induced deformations are small and return to their original positions if the force is removed; that is, domain walls remain trapped in a potential valley (a reversible/pinning regime). In contrast, as the effective force increases, this reversible/pinning regime eventually collapses, and some domain-wall segments begin to exhibit creep, a thermally assisted sluggish motion overcoming potential barriers. Relatively large deformations are thus induced and are no longer reversible (an irreversible/creep regime). This universal behaviour is analogous to that in elastic and plastic regimes in ordered solids; therefore, the electronic media described are sometimes viewed as elastic objects.

Magnetic skyrmions[6–12], spin-swirling topological objects of 10–100 nm in diameter, constitute a new member of elastic objects in electronic systems. Whereas the skyrmion is a pancake-like entity in an ultrathin film, it forms a cylindrical structure in a bulk sample[13] (Fig. 1a), similarly to a flux-line in superconductors. In particular, in bulk chiral magnets, such as MnSi and $(Fe_{1-x}Co_x)Si$, magnetic skyrmion strings are observed as a thermodynamically stable skyrmion-string-lattice (SkS-L) phase[8,9] in which skyrmion strings pointing along an external magnetic field are arranged in the form of a close-packed triangular lattice. Notably, it has been found for MnSi that the thermodynamically stable SkS-L begins to move at extremely low electric current density[14,15], on the order of $10^6$ A m$^{-2}$, which is 5–6 orders of magnitude smaller than the value required for current-induced ferromagnetic domain-wall motion[16,17]. Magnetic skyrmions are therefore attracting considerable research attention as a potential candidate for next-generation information carriers[18–20].

In current-drive racetrack-type applications[21] based on isolated skyrmions[22,23], one must manipulate a metastable isolated skyrmion because thermodynamically stable skyrmions have a propensity to condense and entirely fill a sample/device[10,24]; thus, the behaviour of metastable isolated pancake-like skyrmions under current is of practical interest. In this context, ferromagnetic ultrathin-film stacks, such as polycrystalline Pt/Co/Ta and single-crystalline Pt/CoFeB/MgO, have recently been investigated[23], and it has been observed that, whereas almost all skyrmions move uniformly in the single-crystalline Pt/CoFeB/MgO, pinned and moving skyrmions coexist in polycrystalline Pt/Co/Ta. Notably, in the latter case, a moving skyrmion sometimes collides with a pinned one and irreversibly merges into one pinned skyrmion[23]. These observations suggest that under the influence of random pinning potential, current-induced skyrmion dynamics may be more complex than the uniform flow of a fixed number of skyrmions, particularly near the critical current density at which skyrmions begin to move.

The situation may be even more complex in metastable skyrmion strings in a bulk sample. Notably, the diameter of a skyrmion is 10–100 nm, whereas the skyrmion-string length is on the order of the sample thickness (typically, 10–1000 μm). Ordinary bulk crystals at finite temperatures contain a finite density of defects, such as vacancies, as the most thermodynamically stable state, and a defect-free crystal has a higher free energy[25]. Analogously, such a long skyrmion string is unlikely to be intact at finite temperatures but should contain finite topological defects, that is, pairs of the emergent magnetic monopole and antimonopole[26–28], as schematically shown in Fig. 1a,b. Although the stability and/or dynamics of the emergent magnetic (anti)monopoles are likely directly linked with the controllability of the metastable skyrmion strings, the details remain unclear because of the lack of experiments targeting metastable skyrmion strings in a bulk sample.

In this article, we describe the behaviour of the metastable SkS-L in MnSi under pulsed electric current by means of the Hall resistivity, $\rho_{yx}$, which is a sensitive probe for condensed topological skyrmions[29,30], particularly in MnSi[31–34]. We first write a metastable SkS-L in a limited area of the bar-shaped sample by using a local thermal quenching technique[34–36] while keeping the other area in the conical state, which is the thermodynamically most stable at the magnetic field investigated; next, $\rho_{yx}$ is measured at two distant positions before and after in-plane current pulse applications to determine how the pulse affects the written metastable SkS-L. We find that, whereas the written metastable SkS-L is maintained (the reversible/pinning regime) at low current densities (1–2×10$^6$ A m$^{-2}$), it undergoes irreversible topological unwinding at higher current densities (>3×10$^6$ A m$^{-2}$) rather than being transported along the current direction. Moreover, the critical current density above which the topological unwinding appears is larger for a shorter pulse width, reminiscent of viscoelastic characteristics accompanying a pinning-creep transition in domain-wall motion. The smallness of the involved current density and the numerical simulation results suggest that the topological unwinding is initiated by the current-induced depinning of originally segmented skyrmion strings.

**Results**

**Sample configuration.** We targeted the archetypal skyrmion-hosting material MnSi and planned to write a metastable SkS-L domain in the matrix of a thermodynamically stable conical order to detect the translation of the SkS-L, if it occurred, by probing the Hall voltage at a distant position. In planning how to prepare the metastable SkS-L in the area of interest, we noted that the SkS-L can be created as a metastable state by employing rapid cooling (>100 K s$^{-1}$) by way of the thermo-equilibrium SkS-L phase (see Figs. 2a,b)[34,37]. If such thermal quenching is locally implemented, then the metastable SkS-L domain is also expected to be written locally.

Figures 2c,d display the sample configuration devised for this purpose. In addition to the standard six contacts $c_1$–$c_6$, two large electrodes $e_1$ and $e_2$ are attached on the sample surface. The contact resistance of $e_1$ and $e_2$ is relatively high, ≈10–15 Ω (for comparison, it is less than 0.1 Ω for the current electrodes $c_1$ and $c_2$), to facilitate local Joule heating; the heating is followed by rapid cooling after the pulse cessation. When a single electric pulse of an appropriate magnitude and width is applied to the $e_1$–$c_2$ pair (or the $e_2$–$c_2$ pair), for example, at 15 K and 0.249 T, the SkS-L-to-conical transition line is quickly crossed during such rapid cooling (Fig. 2a), thus leading to the writing of a metastable SkS-L domain around the electrode $e_1$ (or $e_2$) only. For convenience, we label the area around the electrode $e_1$ and $e_2$ as areas 1 and 2, respectively, and refer to the electric-heating pulse as a writing pulse.

**Local writing of the metastable SkS-L.** The conjectured operation was substantiated by simultaneously probing the Hall voltages near the electrodes $e_1$ and $e_2$. To this end, we measured $\rho_{yx}$–$H$ profiles at the $c_3$–$c_4$ and $c_5$–$c_6$ pairs at 15 K; these are shown in Figs. 2e,f, respectively. When no writing pulse was applied (the black curves), the $c_3$–$c_4$ and $c_5$–$c_6$ pairs exhibited nearly the same profiles, thus reflecting the homogeneous conical order as the initial state. In contrast, after a single writing pulse of 110 mA and 15 ms was applied to the electrode $e_1$ at 0.249 T (for the details of the pulse-magnitude dependence, see Supplementary Fig. 1), an enhanced $\rho_{yx}$ was observed for the $c_3$–$c_4$ pair (Fig. 2e), whereas such an enhancement was not discerned for the $c_5$–$c_6$ pair (Fig. 2f); furthermore, when the magnetic field was subsequently increased (the red curves) or decreased (the green curves) from 0.249 T, $\rho_{yx}$ at the $c_3$–$c_4$ pair remained enhanced and then decreased to the values corresponding to the case in which no writing pulse was applied. This additional Hall signal, $\Delta\rho_{yx}$ ($|\Delta\rho_{yx}| \approx 24$–25 n$\Omega$ cm), known as the topological Hall effect[29,30], is a hallmark of condensed topological skyrmions[31–34], thus demonstrating successful writing of the metastable SkS-L domain in the area 1. We also confirmed that when the writing pulse was applied to the electrode $e_2$, an enhanced $\Delta\rho_{yx}$ of a similar magnitude appeared for the $c_5$–$c_6$ pair, whereas $|\Delta\rho_{yx}| \leq 2$ n$\Omega$ cm for the $c_3$–$c_4$ pair (see Supplementary Fig. 2).

**Current application to the metastable SkS-L.** Having established a method to create metastable SkS-L in an area-selective manner, we were able to address the type of dynamics that emerges from the metastable SkS-L under current applications. After writing a metastable SkS-L in the area 1, we applied positive in-plane current pulses (that is, current flowing from $c_1$ to $c_2$) of various magnitudes with a fixed pulse width, 25 ms, and then measured $\rho_{yx}$ at the $c_3$–$c_4$ and $c_5$–$c_6$ pairs simultaneously to observe the effects of the pulse application.

Figures 3a,b display the topological Hall signal, $\Delta\rho_{yx}$, at the $c_3$–$c_4$ and $c_5$–$c_6$ pairs versus in-plane pulse numbers applied, respectively, for several current magnitudes at 15 K. Here, three aspects can be highlighted. First, at the lowest current density, $+1.0\times10^6$ A m$^{-2}$, the topological Hall signal at the $c_3$–$c_4$ pair remains constant even after 3,000 pulses are applied (Fig. 3a). This finding suggests that the weak current application introduces no appreciable changes to the written metastable SkS-L; that is, at this current density, the metastable SkS-L remains in the reversible/pinning regime. Nevertheless, this regime clearly collapses at higher current densities, and the topological Hall signal (equivalently, the density of skyrmion strings) at the $c_3$–$c_4$ pair decreases by the repetitive pulse applications. Second, in this irreversible regime, the decrease in the topological Hall signal occurs primarily at the early stage of the repetitive pulse applications; the decrease then becomes more moderate as the pulse number increased (Fig. 3a). Third, no appreciable signal is transferred to the Hall signal at the $c_5$–$c_6$ pair (Fig. 3b), even after $\Delta\rho_{yx}$ at the $c_3$–$c_4$ pair largely disappears. The second and third aspects are not compatible with the simplest scenario that the metastable SkS-L domain may move as a whole along the current direction. In such a case, the $\Delta\rho_{yx}$ signal at the $c_3$–$c_4$ pair would remain constant as long as the current-induced shift of the metastable SkS-L were small, and the signal would eventually be transferred to the Hall signal at the $c_5$–$c_6$ pair; these are not the case in the experiments. We also performed the experiments with negative in-plane current pulses (current flowing from $c_2$ to $c_1$), but essentially the same results were obtained (Supplementary Fig. 3).

To gain more insight into the decrease in the topological Hall signal, we performed similar experiments but with the alternating application of positive and negative in-plane pulses. In this pulse sequence, the sum of the total effective force applied (including its sign) is zero by definition; thus, the net shift of the metastable SkS-L is not expected, even if each current pulse were to be accompanied by its finite shift. For comparison, the results are plotted in Fig. 3a, labeled with the current magnitude, +/–8.2 and +/–3.4 (×$10^6$ A $m^{-2}$). Remarkably, these pulse sequences exhibited a similar (or even larger) decrease in the topological Hall signal. Therefore, the observed $\Delta\rho_{yx}$ profiles at the $c_3$–$c_4$ pair cannot be a consequence of the translation of the metastable SkS-L. Alternatively, its destruction can account for this observation: the metastable SkS-L undergoes current-induced topological unwinding into a non-topological spin texture, which is probably the thermodynamically most stable conical state.

Regarding the origin of the topological unwinding, we can rule out a Joule-heating-assisted mechanism. We estimated the increase of the local temperature in the area 1 by measuring the voltage at the $c_3$–$c_4$ pair during the pulse application and found a temperature increase of only 0.2 K (specifically, 15.0 → 15.2 K) under 8.2×$10^6$ A $m^{-2}$ and 125 ms (the maximum and longest current density applied in this study: for details, see Supplementary Fig. 4). Moreover, the estimated lifetime of the metastable SkS-L is beyond $10^{15}$ years at 15 K (ref. 34); thus, the slight temperature increase is expected to play a negligible role in the observed topological unwinding. The appreciable difference of the data between +/–8.2 and +8.2×$10^6$ A $m^{-2}$ (Fig. 3a) also cannot be ascribed to Joule heating, which should be symmetric with respect to the current polarity, at least, for Ohmic contacts (this is the case in the experiments; see Supplementary Fig. 4).

When considering the destruction of skyrmion strings from the perspective of topology, pair creation of the emergent magnetic monopole-antimonopole (Fig. 1a,b) and their subsequent unbinding motion are expected to be involved[26–28], analogous to the nucleation and subsequent growth in ordinary first-order phase transitions. To gain insight into which process is more relevant to the current-induced topological unwinding, it is helpful to consider the magnitude of the involved current density. Both a theoretical paper[28] and our order-of-magnitude estimates (Supplementary Note 1) predict that the monopole-antimonopole pair creation requires a current density of ~$10^{12}$ A $m^{-2}$, orders of magnitude larger than the value used in the present experiments, that is, $10^6$–$10^7$ A $m^{-2}$.

Therefore, we conclude that the monopole-antimonopole pairs preexist in the quenched SkS-L and hence that their current-induced unbinding motion plays a key role in the topological unwinding. Meanwhile, $10^6$ A m$^{-2}$ is the order on which the skyrmion strings start to move [15], thus suggesting that the topological unwinding is also related with the creep of the skyrmion strings. In fact, the current-induced topological unwinding at 10 K was found to be less pronounced than that observed at 15 K (Supplementary Fig. 5), in agreement with the notion of creep, which is innately thermally assisted[38–44].

**Numerical simulations.** To determine how the creep of the skyrmion string and the unbinding motion of the preexisting monopole-antimonopole pairs are correlated with each other in the context of the current-induced topological unwinding, we performed micromagnetic simulations of the chiral magnetic system including randomly distributed pinning sites (for details, see the Methods section). We found that under certain numerical conditions, a segmented skyrmion string in a ferromagnetic background can be metastable at zero or weak currents (Fig. 4a). Nevertheless, when the applied current exceeds a specific threshold value, the segmented skyrmion strings start to move; remarkably, this process is also accompanied by the motion of (anti)monopole such that the skyrmion string shortens, as shown in Fig. 4b-d (see also Supplementary Movie 1). As a result, the metastable segmented skyrmion string quickly disappears without a long-distance creep (Fig. 4d). This numerical result establishes a clear example of topological unwinding being initiated by the onset of the creep motion of metastable segmented skyrmion strings, corroborating the experimental observations.

**Kinetic nature of the regime transition.** We then focused on the kinetic nature of the transition between the reversible/pinning and topological unwinding regimes. In the case of domain walls, an irreversible creep motion occurs more frequently under a larger d.c. force[38–44]. When a pulsed or a.c. external force is applied, however, the pulse width or frequency dictates the onset of creep motion as well: at a given effective-force magnitude, domain-wall segments remain pinned and behave like an elastic solid on short timescales (or at high frequencies), whereas they begin to creep like a viscous fluid over longer time scales (or at low frequencies)[1,45–48]. Such a dual nature depending on timescale or frequency may be described as a viscoelastic characteristic and can be qualitatively accounted for by considering that deformations exceeding the reversible/pinning regime cannot be induced instantaneously. Although this feature appears to be involved in a wide class of pinned electronic elastic objects, such a viscoelastic nature has not been explored in the skyrmion strings to date.

To address this issue, we investigated the decrease in the topological Hall signal for various pulse widths, $t_p$, with a sufficiently large pulse magnitude, $+8.2 \times 10^6$ A m$^{-2}$ (Fig. 5a). As expected, relative to the case of long pulse widths, such as $t_p \geq 25$ ms, much less topological unwinding was observed for the shortest pulse width, 5 ms (for the apparent convergence of the $\Delta\rho_{yx}$ profiles for $t_p \geq 25$ ms, see Supplementary Note 2). However, it should be emphasized that the low degree of topological unwinding for $t_p = 5$ ms occurred not because the integrated pulse width (that is, the pulse number multiplied by each $t_p$) was the shortest: As illustrated in Fig. 5b, even if the $\Delta\rho_{yx}$ profiles are compared at a given integrated pulse width, the data for $t_p = 5$ ms still exhibits the least topological unwinding. This finding suggests that the metastable SkS-L tends to remain in the reversible/pinning regime for a short-pulse application, whereas it is more likely to enter the topological unwinding regime for a longer-pulse application. This timescale-dependent dynamic transition is analogous to the viscoelastic characteristics accompanying the pinning-creep transition of domain-wall motion[1,45–48], again highlighting the key role of the pinning-creep transition of the segmented skyrmion strings at the onset of topological unwinding.

**Pulse width vs. critical current density.** The timescale-dependent regime-transition of the metastable SkS-L suggests that the critical current density above which the topological unwinding manifests itself is a function of the pulse width. In estimating the critical current density for various pulse widths, we introduce the quantity $F$ ($:0 \leq F \leq 1$):

$$F(n) \equiv 1 - \Delta\rho_{yx}(n)/\Delta\rho_{yx}(0), \qquad (1)$$

where $\Delta\rho_{yx}(n)$ is a value measured after the positive in-plane current pulse is applied $n$ times and $\Delta\rho_{yx}(0)$ denotes a value after applying the writing pulse; thus, $F(n)$ represents the unwound fraction of the metastable SkS-L after $n$-pulse applications, and $F(n) = 0$ and 1 correspond to the as-written SkS-L and fully unwounded non-topological state, respectively.

Figure 6a summarizes the results at 15 K after 200 pulse applications, $F(n = 200)$, as a function of the current density, $j$, for various pulse widths, $t_p$, showing the clear dependence of $F$ on $j$ and $t_p$. In all cases, the $F(n = 200)$–$j$ profiles (and also the $F(n = 100)$–$j$ profiles: see Supplementary Fig. 6) are well described by a power law with a fixed exponent, $F(n = 200) \sim j^{\alpha}$ with $\alpha = 3.4$ (for $F(n = 100)$, $\alpha = 3.5$: see also Supplementary Fig. 6). This power-law behaviour suggests that, in a strict sense, the critical current density is infinitesimally small; however, this consequence is consistent with the current understanding of creep motion, in which an infinitesimal force allows for creep motion with a finite probability as long as the temperature is finite[38–44].

Nevertheless, we can conditionally derive a critical current density of a practical sense, above which a finite $F(n)$ becomes experimentally appreciable: in view of the present experimental accuracy, we set this criterion as $F(n) = 0.03$ and label the (practical) critical current density with $j_{c,n}$. Figure 6b displays $j_{c,n = 100}$ and $j_{c,n = 200}$ at 15 K as a function of pulse width; although $j_{c,n = 200}$ is reasonably lower than $j_{c,n = 100}$ at a given pulse width, a clear correlation between pulse width and critical current density can be seen for the both profiles, thus highlighting the viscoelastic characteristics at the onset of topological unwinding. Although under the present definition, the value of $j_c$ depends on the pulse number and pulse width, it can be safely concluded that $j_c$ in the limit of long pulse ($t_p \to \infty$) is on the order of $10^6$ A m$^{-2}$, in good agreement with the value reported for the thermodynamically stable SkS-L under d.c. currents[14,15].

**Power-law analysis of the pulse-width dependence.** The $j_c$–$t_p$ profiles at 15 K shown in Fig. 6b are both described by a power law with nearly the same exponent: $j_c - j_c(t_p \to \infty) \sim t_p^{-1/\delta}$ with $\delta = 0.43 \pm 0.05$. The $F(n)$–$j$ profiles at a lower temperature, 10 K, were also analyzed using power laws, but at relatively large $n$ ($n$ = 200 and 1,000) because of the less pronounced topological unwinding (Supplementary Fig. 6); in this manner, we obtained $\alpha$ = 2.4–2.5 and $\delta$ = 0.50±0.06 at 10 K (Supplementary Fig. 7), to be compared with $\alpha$ = 3.4–3.5 and $\delta$ = 0.43±0.05 at 15 K. Thus, unlike the exponent $\alpha$, the exponent $\delta$ appears to vary only weakly in the temperature range of 10–15 K. Although there are currently no numerical results to be compared, the value of $\delta$ would be a potential basis to test the microscopic model for the current-induced dynamics of the metastable SkS-L. At a minimum, the microscopic models for the depinning kinetics have been discussed in charge density waves by comparing the experimentally and theoretically derived $\delta$ values[49–52].

**Discussion**

We expect that the innately segmented skyrmion strings are likely to be relevant also to the current-induced dynamics of the thermodynamically stable SkS-L. Nevertheless, the consequence of the creep motion is expected to be crucially different between the metastable and thermodynamically stable cases: In a metastable SkS-L, the monopoles and antimonopoles move to eliminate the skyrmion string unless they are pinned, eventually lowering the system's total free energy. In a thermodynamically stable SkS-L, by contrast, the topological unwinding is not triggered because the breakdown of the thermodynamically stable state inevitably increases the free energy. The current-induced dynamics of the thermodynamically stable SkS-L are therefore expected to involve the steady flow of the monopoles and antimonopoles, which may be accompanied by fluctuations in the emergent electric field[27,28].

We have also shown that the existing emergent magnetic (anti)monopoles are harmful when one tries to drive metastable skyrmions with an electric current; therefore, in the context of the skyrmion application, the device thickness must be chosen so that skyrmion strings (or cylinders) do not contain topological defects. In the thermo-equilibrium SkS-L phase, the order of the expected segment length is given as $a \times \exp(\Delta_{\text{MP-AMP}}/k_B T)$, where $a$ is the lattice constant of the considered material and $\Delta_{MP-AMP}$ is the pair-creation energy of the monopole and antimonopole. According to the numerical results[27], $\Delta_{MP-AMP}$ at the lowest temperature is $\approx 6J$ for a simple cubic lattice, where $J$ is the magnetic exchange energy and approximately equals to the magnetic transition temperature[53]; hence, for MnSi, $\Delta_{MP-AMP}/k_B$ $\approx 200$ K at low temperatures. Thus, in the thermo-equilibrium SkS-L phase ($\approx 27$ K), the expected segment length is ~700 nm (or shorter, given that $\Delta_{MP-AMP}$ may decrease at high temperatures). Because our sample thickness is $\approx 100$ μm, one can expect that each quenched skyrmion string contains monopole-antimonopole pairs on the order of 100, provided that the topological defect density is frozen at the value of the thermo-equilibrium phase. The direct observation of a segmented skyrmion string remains a challenging issue, which may nevertheless provide useful information when designing a topological-defect-free skyrmion device.

**Methods**

**Sample preparation and setup.** A single crystal of MnSi was grown via the Czochralski method. The sample was cut and polished to a size of $2.5 \times 1.1 \times 0.1$ mm$^3$, with the largest surface normal to the <100> axis. The resistivity ratio $\rho(300\ \text{K})/\rho(4.2\ \text{K})$ was $\approx 52$. Gold current leads of 0.3 mm ϕ were attached to the sample and fixed with indium. The electrodes, $e_1$ and $e_2$, were constructed using carbon paste to achieve a high contact resistance, $\approx 10$–15 Ω. The sample was mounted on a sapphire substrate in contact with a heat bath and fixed with varnish.

**Transport measurements.** The Hall resistivity, $\rho_{yx}$, was measured at 33 Hz with a low a.c. current excitation ($\approx 7.97 \times 10^4$ A m$^{-2}$) under a magnetic field parallel to the <100> axis using lock-in amplifiers (Stanford Research Systems, SR830) equipped with a transformer preamplifier (Stanford Research Systems, SR554). The Hall resistivity values presented here are antisymmetrized between the positive and negative magnetic fields. The pulse currents that were used to write the metastable SkS-L or to exert an effective force on the written metastable SkS-L were generated by a function generator (NF Corporation, WF1947) connected to a bipolar amplifier (NF Corporation, 4502A). Before applying the writing pulse to the electrode $e_1$ at +0.249 T (or –0.249 T), we applied a field of 1 T (or –1 T) to enter the ferromagnetic state and erase any residue of the metastable SkS-L in the previous measurements. When measuring $\Delta\rho_{yx}$ as a function of the in-plane pulse number, we waited 1 min after the last pulse was applied to ensure that the sample temperature was sufficiently equilibrated with the sample-holder temperature. When multiple pulses were applied between measurement points, we set the interval between successive two pulses to be 6 s.

**Micromagnetic simulation.** The simulation was performed for a simple cubic lattice consisting of 60×30×100 magnetic moments with an open boundary condition in the $z$ direction and periodic boundary conditions in the $x$ and $y$ directions. We considered the following model Hamiltonian:

$$H = -J \sum_r \mathbf{n}_r \cdot \left(\mathbf{n}_{r+e_x} + \mathbf{n}_{r+e_y} + \mathbf{n}_{r+e_z}\right)$$
$$+ D \sum_r (\mathbf{n}_r \times \mathbf{n}_{r+e_x} \cdot \mathbf{e}_x + \mathbf{n}_r \times \mathbf{n}_{r+e_y} \cdot \mathbf{e}_y + \mathbf{n}_r \times \mathbf{n}_{r+e_z} \cdot \mathbf{e}_z)$$
$$- h \sum_r n_{z,r} - K_{\text{imp}} \sum_{r \in \Lambda} (n_{z,r})^2, \qquad (2)$$

where $J$ is the exchange interaction, $D$ is the Dzyaloshinskii-Moriya interaction energy, $h_z$ is the magnetic field along the $z$ direction, $\mathbf{e}_x$ (or $\mathbf{e}_y$, $\mathbf{e}_z$) is the unit vector that connects with the nearest neighbour site along the $x$ (or $y$, $z$) direction, $\mathbf{n}_r$ is the unit vector of the local magnetic moment at site $r$, and $n_{z,r}$ is the $z$ component of $\mathbf{n}_r$. The last term represents the impurity in the model: the easy axis anisotropy $K_{\text{imp}}$ was introduced at randomly selected sites, and $\Lambda$ is the set of random numbers. In simulating the current-induced dynamics of the skyrmion strings at zero temperature, we inserted in the Hamiltonian into the following Landau-Lifshitz-Gilbert equation:

$$\frac{d\mathbf{n}_r}{dt} = -\gamma \frac{dH}{d\mathbf{n}_r} \times \mathbf{n}_r + \alpha \mathbf{n}_r \times \frac{d\mathbf{n}_r}{dt} - (\mathbf{j} \cdot \nabla)\mathbf{n}_r + \beta[\mathbf{n}_r \times (\mathbf{j} \cdot \nabla)\mathbf{n}_r], \qquad (3)$$

where $\mathbf{j}$ represents the (spin-polarized) electric current density. The units of non-dimensional time $t$ and current $j = |\boldsymbol{j}|$ are $1/(\gamma J)$ and $2e\gamma J/pa^2$ ($p$: polarization of the magnet), respectively. We chose the following parameter set {$J = 1.0$, $D = 0.2$, $K_{\mathrm{imp}} = 0.2$, $h = 0.018$, $\alpha = \beta = 0.04$, $j = 0.04$}. The density of the random impurity was set as 10 %.

**Data availability.** The data that support the findings of this study are available from the corresponding author on request.

## References

1. Kleemann, W. Universal Domain Wall Dynamics in Disordered Ferroic Materials. *Annu. Rev. Mater. Res*. **37,** 415–448 (2007).
2. Catalan, G., Seidel, J., Ramesh, R. & Scott, J. F. Domain wall nanoelectronics. *Rev. Mod. Phys*. **84,** 119–156 (2012).
3. Blatter, G., Feigel'man, M. V., Geshkenbein, V. B., Larkin, A. I. & Vinokur, V. M. Vortices in high-temperature superconductors. *Rev. Mod. Phys*. **66,** 1125–1388 (1994).
4. Grüner, G. The dynamics of charge-density waves. *Rev. Mod. Phys*. **60,** 1129–1181 (1988).
5. Grüner, G. The dynamics of spin-density waves. *Rev. Mod. Phys*. **66,** 1–24 (1994).
6. Bogdanov, A. & Yablonskii, D. A. Thermodynamically stable "vortices" in magnetically ordered crystals. The mixed state of magnets. *Sov. Phys. JETP* **68,** 101–103 (1989).
7. Bogdanov, A. & Hubert, A. Thermodynamically stable magnetic vortex states in magnetic crystals. *J. Magn. Magn. Mater*. **138,** 255–269 (1994).
8. Mühlbauer, S. *et al*. Skyrmion Lattice in a Chiral Magnet. *Science* **323,** 915–919 (2009).
9. Münzer, W. *et al*. Skyrmion lattice in the doped semiconductor Fe$_{1-x}$Co$_x$Si. *Phys. Rev. B* **81,** 041203 (2010).
10. Yu, X. Z. *et al*. Real-space observation of a two-dimensional skyrmion crystal. *Nature* **465,** 901–904 (2010).

11. Pfleiderer, C. *et al*. Skyrmion lattices in metallic and semiconducting B20 transition metal compounds. *J. Phys.: Condens. Matter* **22**, 164207 (2010).

12. Nagaosa, N. & Tokura, Y. Topological properties and dynamics of magnetic skyrmions. *Nat. Nanotech.* **8,** 899–911 (2013).

13. Park, H. S. *et al*. Observation of the magnetic flux and three-dimensional structure of skyrmion lattices by electron holography. *Nat. Nanotech.* **9,** 337–342 (2014).

14. Jonietz, F. *et al*. Spin Transfer Torques in MnSi at Ultralow Current Densities. *Science* **330,** 1648–1651 (2010).

15. Schulz, T. *et al*. Emergent electrodynamics of skyrmions in a chiral magnet. *Nat. Phys.* **8,** 301–304 (2012).

16. Grollier, J. *et al*., Switching a spin valve back and forth by current-induced domain wall motion. *Appl. Phys. Lett*. **83,** 509–511 (2003).

17. Tsoi, M., Fontana, R. E. & Parkin, S. S. P. Magnetic domain wall motion triggered by an electric current. *Appl. Phys. Lett*. **83,** 2617–2619 (2003).

18. Fert, A., Cros, V. & Sampaio, J. Skyrmions on the track. *Nat. Nanotech.* **8,** 152–156 (2013).

19. Sampaio, J., Cros, V., Rohart, S., Thiaville, A. & Fert, A. Nucleation, stability and current-induced motion of isolated magnetic skyrmions in nanostructures. *Nat. Nanotech.* **8,** 839–844 (2013).

20. Iwasaki, J., Mochizuki, M. & Nagaosa, N. Current-induced skyrmion dynamics in constricted geometries. *Nat. Nanotech.* **8,** 742–747 (2013).

21. Parkin, S. S. P., Hayashi, M. & Thomas, L. Magnetic Domain-Wall Racetrack Memory. *Science* **320,** 190–194 (2008).

22. Jiang, W. *et al*. Blowing magnetic skyrmion bubbles. *Science* **349,** 283–286 (2015).

23. Woo, S. *et al*. Observation of room-temperature magnetic skyrmions and their current-driven dynamics in ultrathin metallic ferromagnets. *Nat. Mat.* **15,** 501–506 (2016).

24. Romming, N. *et al*. Writing and deleting single magnetic skyrmions. *Science* **341,** 636–639 (2013)

25. Porter, D. A., Easterling, K. E. & Sherif, M. Y. *Phase Transformations in Metals and Alloys* (CRC Press, 2008).

26. Milde, P. *et al*. Unwinding of a Skyrmion Lattice by Magnetic Monopoles. *Science* **340,** 1076–1080 (2013).

27. Schütte, C. & Rosch, A. Dynamics and energetics of emergent magnetic monopoles in chiral magnets. *Phys. Rev. B* **90,** 174432 (2014).

28. Lin, S-Z. & Saxena, A. Dynamics of Dirac strings and monopolelike excitations in chiral magnets under a current drive. *Phys. Rev. B* **93,** 060401 (2016).

29. Ye, J. *et al*. Berry Phase Theory of the Anomalous Hall Effect: Application to Colossal Magnetoresistance Manganites. *Phys. Rev. Lett.* **83,** 3737–3740 (1999).

30. Bruno, P., Dugaev, V.K. & Taillefumier, M. Topological Hall Effect and Berry Phase in Magnetic Nanostructures. *Phys. Rev. Lett.* **93,** 096806 (2004).

31. Lee, M., Kang, W., Onose, Y., Tokura, Y. & Ong, N. P. Unusual Hall Effect Anomaly in MnSi under Pressure. *Phys. Rev. Lett.* **102,** 186601 (2009).

32. Neubauer, A. *et al*. Topological Hall Effect in the *A* Phase of MnSi. *Phys. Rev. Lett.* **102,** 186602 (2009).

33. Liang, D., DeGrave, J. P., Stolt, M. J., Tokura, Y. & Jin, S. Current-driven dynamics of skyrmions stabilized in MnSi nanowires revealed by topological Hall effect. *Nat. Commun.* **6,** 8217 (2015).

34. Oike, H. *et al*. Interplay between topological and thermodynamic stability in a metastable magnetic skyrmion lattice. *Nat. Phys.* **12,** 62–66 (2016).

35. Oike, H. *et al*. Phase-change memory function of correlated electrons in organic conductors. *Phys. Rev. B* **91,** 041101 (2015).

36. Kagawa, F. & Oike, H. Quenching of charge and spin degrees of freedom in condensed matter. *Adv. Mater.* **29,** 1601979 (2017).

37. Nakajima, T. *et al*. Skyrmion-Lattice Structural Transition in MnSi. *Sci. Adv.* **3,** e1602562 (2017).

38. Miller, R. C. & Savage, A. Velocity of Sidewise 180 Domain-Wall Motion in $BaTiO_3$ as a Function of Applied Electric Field. *Phys. Rev.* **112,** 755–762 (1958).

39. Chauve, P., Giamarchi, T. & Le Doussal, P. Creep and creep in disordered media. *Phys. Rev. B* **62**, 6241–6267 (2000).

40. Kolton, A. B., Rosso, A., Giamarchi, T. & Krauth, W. Creep dynamics of elastic manifolds via exact transition pathways. *Phys. Rev. B* **79,** 184207 (2009).

41. Shin, Y.-H., Grinberg, I., Chen, I-W. & Rappe, A. M. Nucleation and growth mechanism of ferroelectric domain-wall motion. *Nature* **449**, 881–884 (2007).

42. Lemerle, S. *et al*. Domain Wall Creep in an Ising Ultrathin Magnetic Film. *Phys. Rev. Lett.* **80,** 849–852 (1998).

43. Tybell, T., Paruch, P., Giamarchi, T. & Triscone, J.-M. Domain Wall Creep in Epitaxial Ferroelectric Pb($Zr_{0.2}Ti_{0.8}$)$O_3$ Thin Films. *Phys. Rev. Lett.* **89,** 097601 (2002).

44. Paruch, P., Giamarchi, T., Tybell, T. & Triscone, J.-M. Nanoscale studies of domain wall motion in epitaxial ferroelectric thin films. *J. Appl. Phys.* **110,** 051608 (2006).

45. Braun, Th., Kleemann, W., Dec, J. & Thomas, P. A. Creep and Relaxation Dynamics of Domain Walls in Periodically Poled $KTiOPO_4$. *Phys. Rev. Lett.* **94,** 117601 (2005).

46. Kagawa, F., Hatahara, K., Horiuchi, S., & Tokura, Y. Domain-wall dynamics coupled to proton motion in a hydrogen-bonded organic ferroelectric. *Phys. Rev. B* **85,** 220101 (2012).

47. Kleemann, W. et al. Modes of Periodic Domain Wall Motion in Ultrathin Ferromagnetic Layers. *Phys. Rev. Lett.* **99,** 097203 (2007).

48. Fedorenko, A. A., Mueller, V. & Stepanow, S. Dielectric response due to stochastic motion of pinned domain walls. *Phys. Rev. B* **70,** 224104 (2004).

49. Mihaly, L., Chen, Ting. & Grüner, G. Switching, hysteresis and time delay in charge-density-wave conduction. *Solid State Commun*. **61,** 751–753 (1987).

50. Inui, M., Hall, R. P., Doniach, D. & Zettl, A. Phase slips and switching in charge-density-wave transport. *Phys. Rev. B* **38,** 13047–13060 (1988).

51. Strogatz, S. H.& Westervelt, R. M. Predicted power laws for delayed switching of charge-density waves. *Phys. Rev. B* **40,** 10501–10508 (1989).

52. Ogawa, N & Miyano, K. Optical investigation of the origin of switching conduction in charge-density waves. *Phys. Rev. B* **70,** 075111 (2004).

53. Buhrandt, S. & Fritz, L. Skyrmion lattice phase in three-dimensional chiral magnets from Monte Carlo simulations. *Phys. Rev. B* **88,** 195137 (2013).

**Acknowledgements**   We thank S. Hoshino, K. Shibata, G. Tatara, N. Ogawa and M. Ikeda for fruitful discussions. This work was partially supported by CREST, JST (grant no. JPMJCR16F1) and KAKENHI (grant nos. 25220709, 26103006, 15H03553, 15K05192 and 15H05459).

**Author Contributions**   F.K. performed all the experiments and analyzed the data. A.K. grew the single crystals used for the study. Y.O. cut and polished the sample. F.K. and H.O. planned the project. F.K. wrote the manuscript. W.K. and N.N. performed numerical simulations. All authors discussed the results and commented on the manuscript.

**Additional Information**   Supplementary Information is available in the online version of the paper. Correspondence and requests for materials should be addressed to F.K.

**Competing financial interests**   The authors declare no competing financial interests.

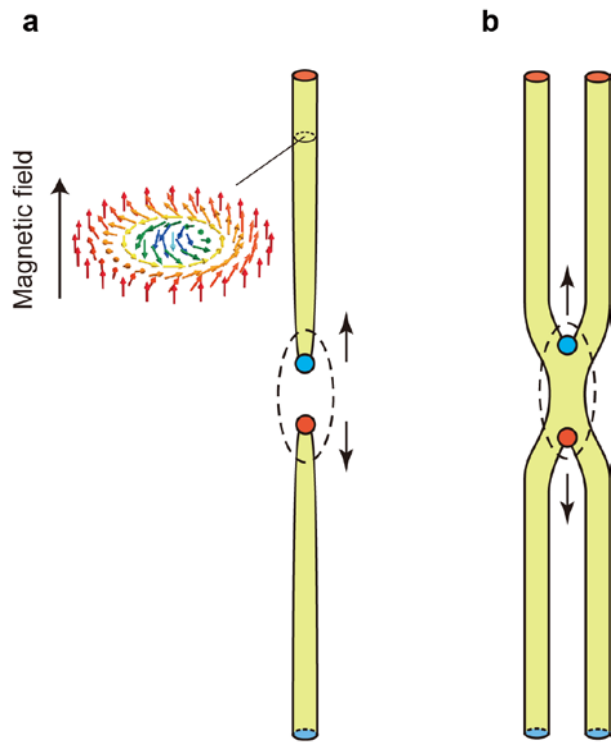

**Figure 1 │ Schematics of the emergent magnetic monopole and antimonopole.**
**a,b,** Monopoles/antimonopoles resulting from the pinching-off of a skyrmion string (**a**) and partial merging of two neighbouring skyrmion strings (**b**). Coloured circles represent emergent magnetic monopoles/antimonopoles. Arrows indicate expected motions of the monopoles/antimonopoles when the topological unwinding progresses.

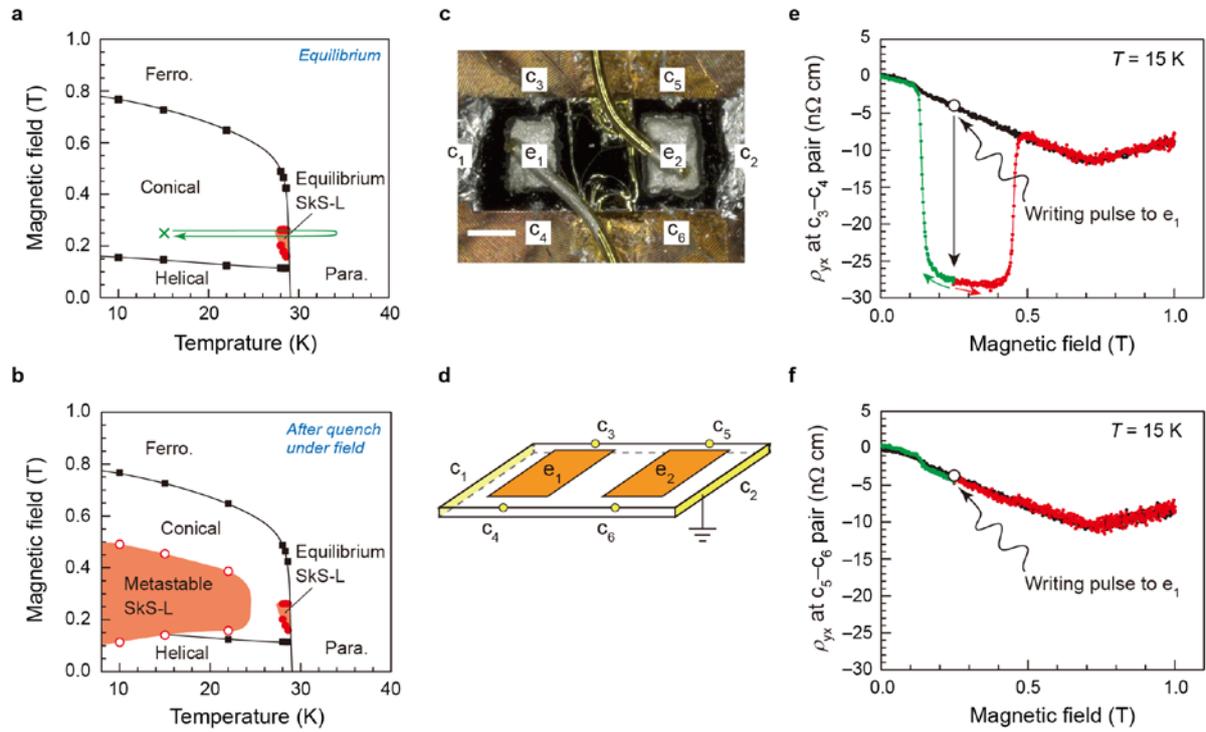

**Figure 2 │ Area-selective writing of the metastable SkS-L in MnSi. a,b,** Magnetic state diagrams of MnSi used in this study under equilibrium (**a**) and quenched conditions (**b**). Thermal quenching was performed by applying the writing pulse to the $e_1$–$c_2$ pair (see **c,d**) at 0.249 T; the thermal history that follows the pulse application is schematically shown with the arrow in **a**. The overall features of the phase diagram were reconstructed by referring to the literature[8,34], and the magnetic-phase boundaries in the present sample were determined by using a previously reported procedure[34]. **c,d,** Photograph (**c**) and schematic diagram (**d**) of the sample configuration. The scale bar is 500 μm. **e,f,** The magnetic-field dependence of $\rho_{yx}$ at 15 K, simultaneously measured at the $c_3$–$c_4$ pair (**e**) and the $c_5$–$c_6$ pair (**f**). Black curves are the data measured before application of the writing pulse (110 mA and 25 ms) to the electrode $e_1$, whereas the red and green ones are those after. The effects of demagnetizing fields are not calibrated.

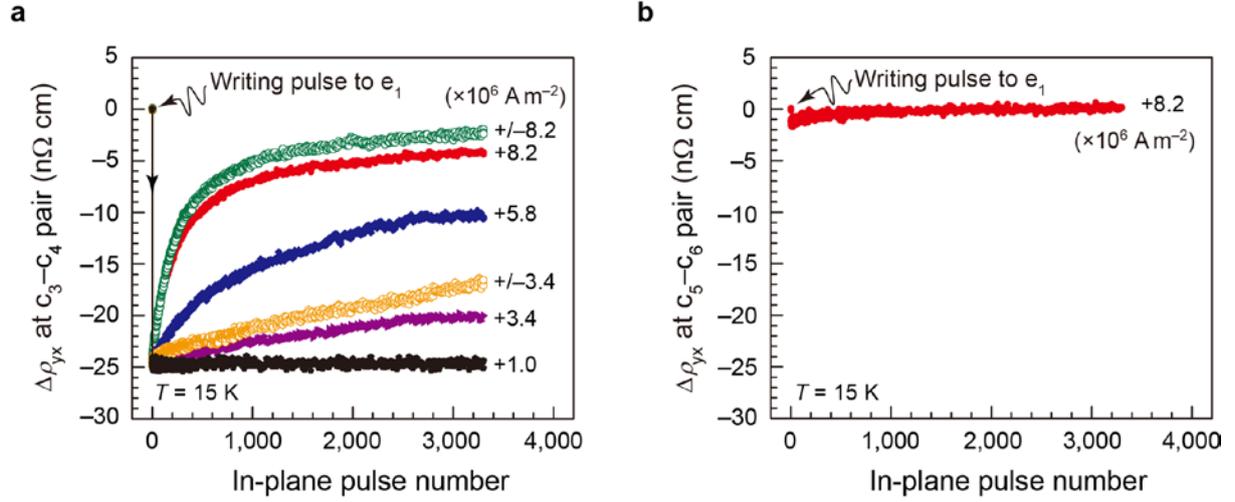

**Figure 3 | Topological unwinding of the metastable SkS-L under current. a,** Variation in $\Delta\rho_{yx}$ at the $c_3$–$c_4$ pair for various pulse magnitudes in response to repeated in-plane pulses. **b,** $\Delta\rho_{yx}$ versus in-plane pulse number at the $c_5$–$c_6$ pair for an in-plane current pulse of $+8.2\times10^6$ A m$^{-2}$. $\Delta\rho_{yx}$, a good measure of the topological Hall signal, is obtained by offsetting the $\rho_{yx}$ value before applying the writing pulse to zero. Labels, such as "8.2", represent the in-plane pulse magnitude used in the measurements, expressed in the unit of $10^6$ A m$^{-2}$. + (closed symbols) denotes the current flowing from $c_1$ to $c_2$ (see Figs. 2c,d), and +/– (open symbols) denotes the pulse sequence in which positive and negative pulses are applied in an alternating manner. Measurements were performed at 15 K and 0.249 T with a fixed pulse width of 25 ms.

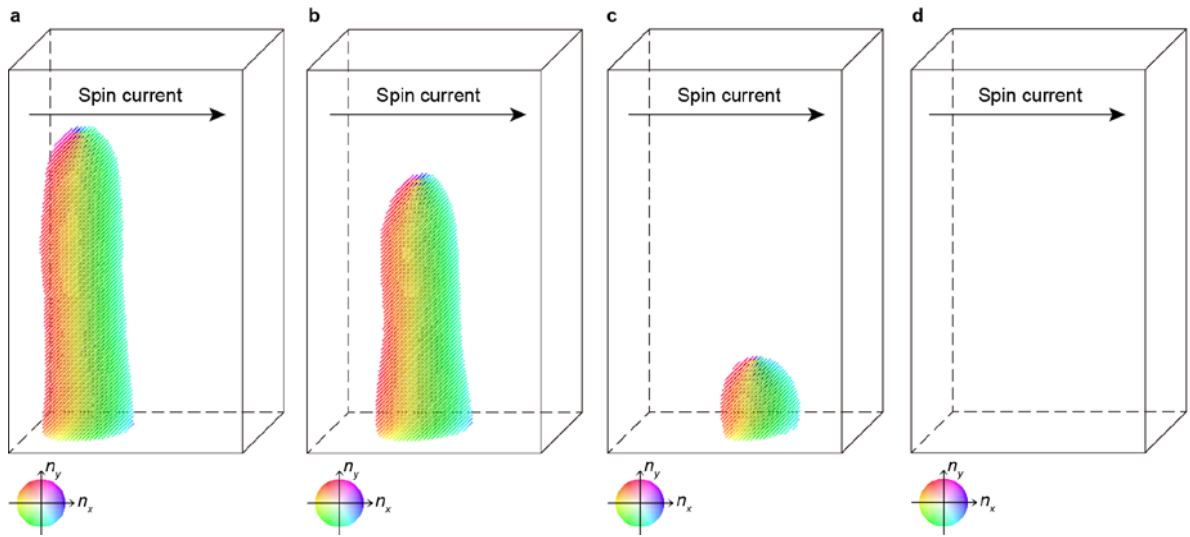

**Figure 4 │ Micromagnetic simulations for a segmented skyrmion string. a,** Initial magnetic state of the simulations, which is metastable under zero current. **b,c,d,** Snapshots of the current-induced dynamics of a segmented skyrmion string under $j =$ 0.04: $t = 250$ (**b**), $t = 500$ (**c**), and $t = 650$ (**d**) in units of $1/(\gamma J)$ (see the Methods section). The movie is available in Supplementary Movie 1. Colour wheels specify the $x$-$y$ plane magnetization direction. The brightness of the color represents the $z$ component of the magnetization; that is, the local magnetizations pointing toward the $z$ direction are displayed as white. For the details of the calculation, see the Methods section.

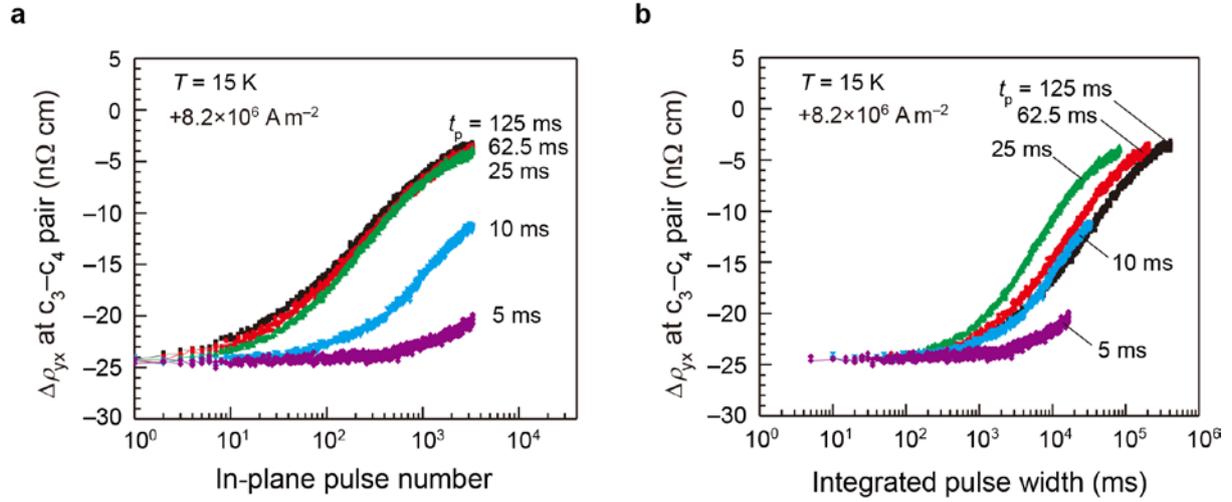

**Figure 5 │Viscoelastic characteristics of the topological unwinding. a,b,** $\Delta\rho_{yx}$ versus in-plane pulse number (**a**) and versus integrated pulse width (**b**) for various pulse width, $t_p$, plotted on a logarithmic scale. The integrated pulse width represents the pulse number multiplied by each pulse width; the data shown in **a,b** are the same data presented in a different manner. Measurements were performed at 15 K and 0.249 T with a fixed pulse magnitude, $+8.2\times10^6$ A m$^{-2}$ (current flowing from $c_1$ to $c_2$: see Figs. 2c,d).

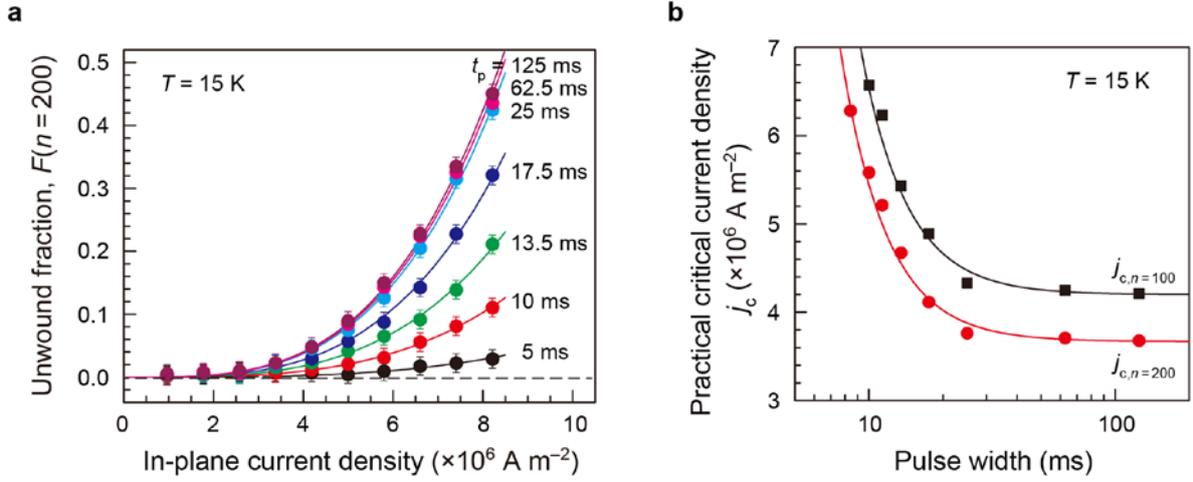

**Figure 6 │ Pulse-width dependence of the critical current density. a,** The topologically unwound fraction of the written metastable SkS-L after the application of 200 in-plane pulses, $F(n = 200)$, with varying current density, $j$, and pulse width, $t_p$. The curves are fits to the power low with a fixed exponent, $F(n = 200) \sim j^{3.4}$. The error bars are derived from the uncertainty about the measured $\rho_{yx}$ values corresponding to ±0.4 nΩ cm. **b,** Practical critical current density versus pulse width. $j_{c,n=100}$ and $j_{c,n=200}$ denote the current density at which $F(n = 100)$ and $F(n = 200)$, respectively, are equal to 0.03 at a given pulse width. Black and red lines are fits to the power law, $j_{c,n} - j_{c,n}(t_p \to \infty) \sim t_p^{-1/\delta}$, with $\delta = 0.43$. Measurements were performed at 15 K and 0.249 T.

## Supplementary note 1: Order-of-magnitude estimates of the current density required for the monopole-antimonopole pair creation

The creation energy of the monopole and antimonopole pair, $\Delta_{\text{MP-AMP}}$, is approximately given as $\approx 6J$ for a simple cubic lattice, where $J$ is the magnetic exchange interaction energy[27]. For the case of MnSi, the magnetic transition temperature ($\approx 1.0J$: ref. 53) is ~30 K; thus, $\Delta_{\text{MP-AMP}}/k_B$ is estimated to be ~200 K.

Below, on the basis of the Thiele equation, we estimate the order-of-magnitude of the charge current density that is required to compensate $\Delta_{\text{MP-AMP}}$. The Thiele equation for a skyrmion under a steady state is given as follows:

$$4\pi \hat{z} \times (\vec{j_s} - \vec{v_d}) + \kappa(\beta \vec{j_s} - \alpha \vec{v_d}) + \vec{F} = 0, \tag{1}$$

where $\vec{j_s}$ represents the spin-current density, $\vec{v_d}$ is the skyrmion drift velocity, $\hat{z}$ is the unit vector along the magnetic field direction, $\alpha$ is the Gilbert damping constant, $\beta$ is the coefficient of the non-adiabatic effect, $\vec{F}$ is the force acting on the skyrmion, and $\kappa$ is a constant on the order of unity: all quantities are dimensionless. For simplicity, we consider a situation in which, under a spin current application, a skyrmion segment collides with neighbouring pinned one at a distance of 20 nm ($\approx$ the skyrmion lattice constant of MnSi) in a quasi-static manner (hence, $\vec{v_d} = 0$). Given that $\beta$ is typically on the order of $10^{-1}$–$10^{-2}$ ($\ll 4\pi$), the Thiele equation reduces to:

$$4\pi \hat{z} \times \vec{j_s} + \vec{F} = 0 \tag{2}$$

In the present problem, $\vec{F}$ is the repulsive force between the two skyrmions. As long as one assumes that the MP-AMP pair creation occurs when $\Delta_{\text{MP-AMP}}/k_B$ (~200 K) is compensated by the work that the external spin current has done against the repulsive force, the required force can be estimated to be on the order of $10^{-13}$ N ($= |\vec{F}|J/a$, where $|\vec{F}|$ is the dimensionless force in the Thiele equation and $a$ is the lattice constant for MnSi, $\approx 4.5$ Å). Thus, the required charge current, $\vec{j_e}$ [A/m$^2$], is estimated to be $10^{12}$ A/m$^2$, where $\vec{j_e}$ is given as $\vec{j_s} \times 2e\gamma J/(pa^2)$ with $p$ (= 0.2) being the polarization of the magnet and $\gamma$ being the gyromagnetic ratio. The estimated value is also consistent with that reported in a theoretical paper[28].

**Supplementary note 2: The apparent convergence of the $\Delta\rho_{yx}$ profiles for $t_p \geq 25$ ms**

In Fig. 5a, the $\Delta\rho_{yx}$ profiles roughly converge for pulse widths longer than 25 ms. This observation has two implications. First, when a single current pulse is considered, the aggregate of skyrmion strings eventually reach a steady state under the current application; consequently, further duration of the single current pulse introduces little change to the aggregate, in which some fraction is already relaxed into a non-topological state at the early stage of the pulse duration. Second, because the topological unwinding nevertheless progresses when the next pulse is applied, the partly relaxed aggregate of skyrmion strings appears to be rearranged during a pulse interval and to result in a skyrmion-string aggregate that is distinct from an aggregate immediately before the pulse cessation; such an updated aggregate can further relax if a subsequent pulse is applied. Nevertheless, we note that these implications remain largely speculative.

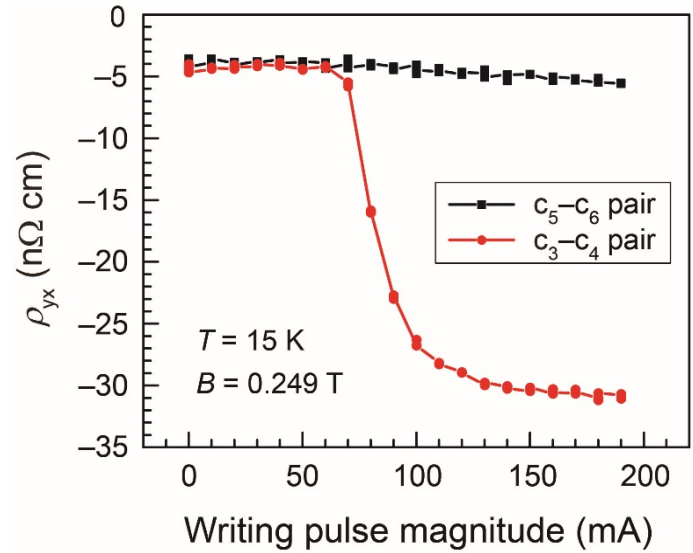

**Supplementary Fig. 1 | $\rho_{yx}$ at the $c_3$–$c_4$ and $c_5$–$c_6$ pairs versus the writing pulse magnitude applied to the electrode $e_1$.** Measurements were performed at 15 K and 0.249 T with a fixed pulse width of 15 ms. When the pulse-current magnitude is weak, $\rho_{yx}$ at the $c_3$–$c_4$ and $c_5$–$c_6$ pairs exhibit nearly the same value, thus indicating the homogeneous conical order. In contrast, when the pulse-current magnitude is sufficiently large, an enhanced $\rho_{yx}$ is observed for the $c_3$–$c_4$ pair, whereas such an enhancement is not discerned for the $c_5$–$c_6$ pair.

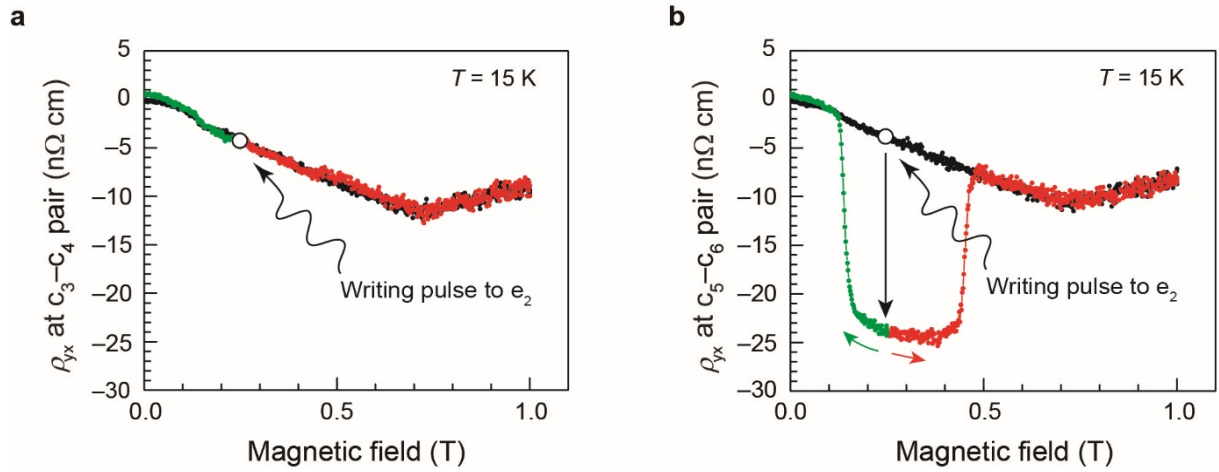

**Supplementary Fig. 2 | Magnetic-field dependence of $\rho_{yx}$ at 15 K, simultaneously measured at the $c_3$–$c_4$ pair (a) and the $c_5$–$c_6$ pair (b).** Black curves are the data measured before application of the writing pulse (150 mA and 15 ms) to the electrode $e_2$, whereas the red and green ones are those after.

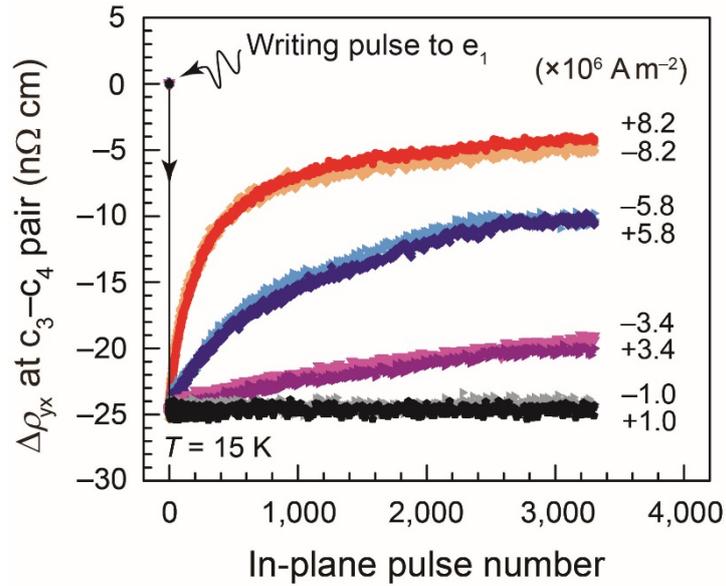

**Supplementary Fig. 3 | Topological unwinding of the metastable SkS-L under positive and negative current-pulse applications.** Labels, such as "8.2", represent the in-plane pulse magnitude used in the measurements, expressed in the unit of $10^6$ A m$^{-2}$. + and − denote current flowing from $c_1$ to $c_2$ and $c_2$ to $c_1$, respectively (see Figs. 2c,d). The data for the positive current are the same with those shown in Fig. 3a. Measurements were performed at 15 K and 0.249 T with a fixed pulse width of 25 ms.

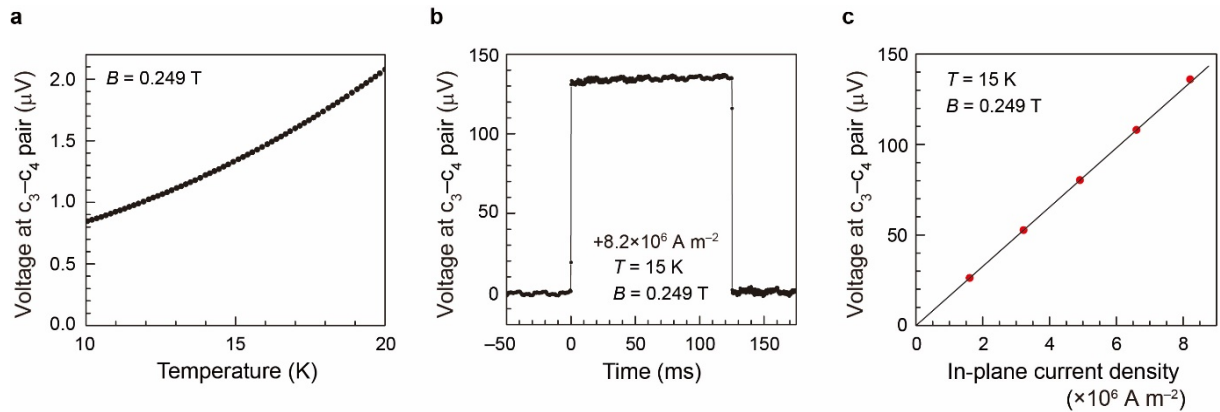

**Supplementary Fig. 4 | Estimation of local temperature increase due to Joule heating. a,** The temperature dependence of voltage at the $c_3$–$c_4$ pair measured with $7.97 \times 10^4$ A m$^{-2}$. Because of a slight misalignment of the contacts $c_3$ and $c_4$, the raw voltage data at the $c_3$–$c_4$ pair are dominated by the longitudinal resistivity, $\rho_{xx}$, unless the measured voltage is antisymmetrized with respect to positive and negative magnetic fields. This voltage-temperature profile can be used as a suitable reference when estimating a local temperature of the area 1. **b,** Time profile of voltage at the $c_3$–$c_4$ pair during the in-plane pulse application of $8.2 \times 10^6$ A m$^{-2}$ with 125 ms width. **c,** Current-voltage characteristics at the $c_3$–$c_4$ pair. The voltages are derived by measuring the voltage-time profile for each pulse magnitude, as shown in **b**. At the largest pulse magnitude, $8.2 \times 10^6$ A m$^{-2}$, the slight non-linearity of ≈2 % is appreciable in the current-voltage characteristics, which can be ascribed to a temperature increase of 0.2 K by referring to the voltage-temperature profile shown in **a**.

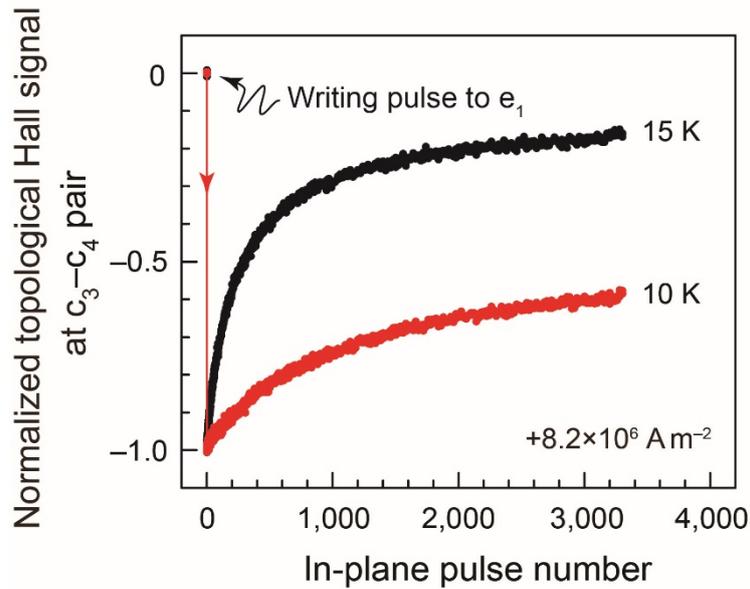

**Supplementary Fig. 5 | Comparison of the degree of topological unwinding between 10 and 15 K.** Because the topological Hall signal is intrinsically temperature dependent (see ref. 34), the normalized topological Hall signals are displayed, thus facilitating a fair comparison of the degree of topological unwinding between different temperatures. Here, −1 and 0 in the vertical axis correspond to the as-written metastable SkS-L and fully non-topological conical state, respectively. The measurements were performed at 0.249 T with a fixed pulse magnitude of $+8.2\times10^6$ A $m^{-2}$ and a pulse width of 25 ms.

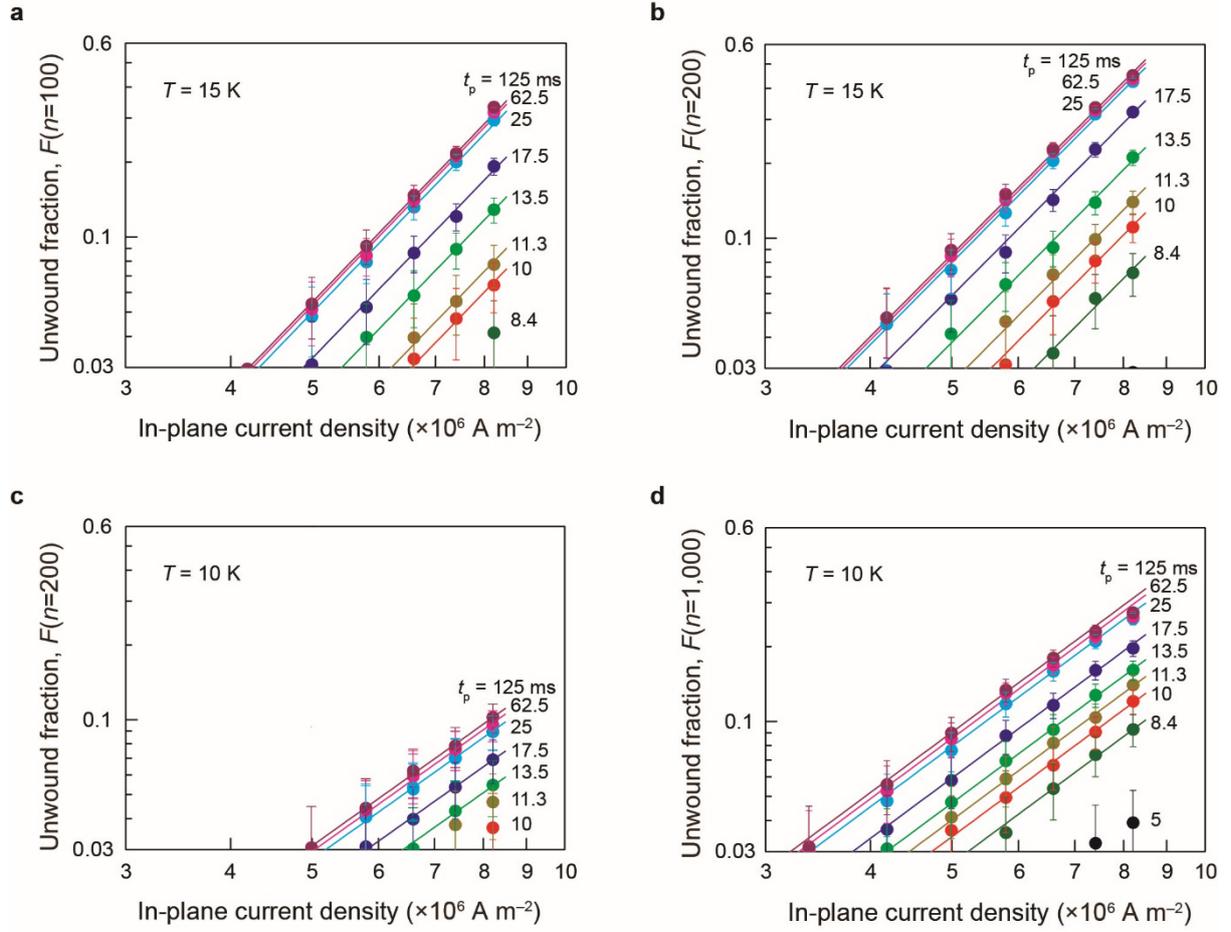

**Supplementary Fig. 6 | Topologically unwound fraction of the metastable SkS-L after application of in-plane pulses $n$ times, $F(n)$, for various values of the current density, $J$, and the pulse width, $t_p$. a,b,** $F(n = 100)$ (**a**) and $F(n = 200)$ (**b**) at 15 K. Lines are fits to the power law, $F \sim J^\alpha$ with $\alpha = 3.5$ for $F(n = 100)$ and $\alpha = 3.4$ for $F(n = 200)$. **c,d,** $F(n = 200)$ (**c**) and $F(n = 1,000)$ (**d**) at 10 K. Lines are fits to the power law, $F \sim J^\alpha$ with $\alpha = 2.4$ for $F(n = 200)$ and $\alpha = 2.5$ for $F(n = 1,000)$. Error bars represent the measurement uncertainty corresponding to ±0.4 nΩ cm. Measurements were performed at 0.249 T.

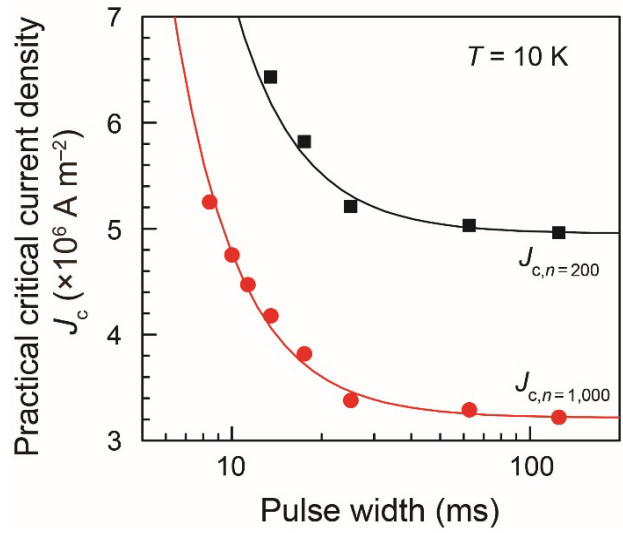

**Supplementary Fig. 7 | Practical critical current density versus pulse width at 10 K.** $J_{c,n=200}$ and $J_{c,n=1,000}$ denote the current density at which $F(n = 200)$ and $F(n = 1,000)$, respectively, are equal to 0.03 at a given pulse width. Black and red lines are fits to the power law, $J_{c,n} - J_{c,n}(t_p \to \infty) \sim t_p^{-1/\delta}$, with $\delta = 0.50$. Measurements were performed at 0.249 T.